\documentclass[aps,12pt,superscriptaddress,amsfonts,amssymb,amsmath]{revtex4}

\usepackage{graphicx}
\usepackage{epsfig}
\usepackage{makeidx}
\usepackage{epstopdf}

\newcommand{\sumsui}{{\sum_{i=1}^n}}
\newcommand{\sumsuh}{{\sum_{h=1}^n}}
\newcommand{\sumsuk}{{\sum_{k=1}^n}}
\newcommand{\sumsua}{{\sum_{\alpha=1}^N}}
\newcommand{\sumsub}{{\sum_{\beta=1}^N}}
\def\QED{\hfill\hbox{\vrule width 4pt height 6pt depth 1.5 pt}}

\begin{document}

\title{Discretized kinetic theory on scale-free networks}

\author{M.L.\ Bertotti \footnote{Email address: marialetizia.bertotti@unibz.it}}

\author{G.\ Modanese \footnote{Email address: giovanni.modanese@unibz.it}}
 
\affiliation{Free University of Bozen-Bolzano, Faculty of Science and Technology, Bolzano, Italy}

\linespread{0.9}

\begin{abstract}

The network of interpersonal connections is one of the possible heterogeneous factors which affect the income distribution emerging from micro-to-macro economic models. 
In this paper we equip our model discussed in \cite{BertottiModanesePhysA,BertottiModaneseEPJB} with a network structure. The model is based on a system of $n$ differential equations of the kinetic discretized-Boltzmann kind. 
The network structure is incorporated in a probabilistic way, through the introduction of a link density $P(\alpha)$ and of correlation coefficients $P(\beta|\alpha)$, which give the conditioned probability that an individual with $\alpha$ links is connected to one with $\beta$ links. 
We study the properties of the equations and give analytical results concerning the existence, normalization and positivity of the solutions.
For a fixed network with $P(\alpha)=c/\alpha^q$, we investigate numerically the dependence of the detailed and marginal equilibrium distributions on the initial conditions and on the exponent $q$.  Our results are compatible with those obtained from the Bouchaud-Mezard model and from agent-based simulations, and provide additional information about the dependence of the individual income on the level of connectivity. 

\end{abstract}

\maketitle

\section{Introduction}

The network of interpersonal connections is an important factor in defining economic interactions of individuals in a society. 
We may suppose that well-connected individuals have more frequent and numerous interactions, with a positive effect on their income. 
Conversely, one might argue that rich individuals can afford more connections than poor ones, because maintaining a link requires some expenses and investment for communication and transportation. 
We may therefore wonder if there is a relation between the link density distribution $P(\alpha)$ and the income distribution, in the sense that, for instance, 
to a larger number $\alpha$ of links corresponds a larger income. 
The assortative or disassortative nature of the network may also have an effect on income distribution. Common sense suggests that in a developed society the correlation is more of a disassortative type, 
namely such that individuals with many links are mainly connected with individuals with few links. 
This could be for instance the case of the economic relationship between the owner of a firm and her employees, or between the owner of a store and his clients. 

Most existing studies of economic interactions on a network do not concern interactions between individuals, but between financial institutions, companies or countries \cite{Fagiolo}. 
When networks of individuals have been considered, they have generally been treated in the context of interacting-agents models. In these models  
simulation algorithms pick individuals at random from an ensemble and let them interact according to certain rules. It is relatively straightforward 
to improve the algorithms by inserting into the list of agents some pre-defined connections which form a certain network \cite{Spagnoli,Chat,Mexican}. 
It is even possible to let the network evolve dynamically, with links changing in dependence on the wealth of the nodes \cite{Abram,Iglesias,Coelho,Hu,Korean}.

Another approach is based on multi-agent linear stochastic equations \cite{Loffredo1,Loffredo2,DiMatAsteHyde}. These have the form $\dot w_i(t)=\eta_i(t) w_i(t) + \sum_{j \neq i} J_{ji} w_j(t) -  \sum_{j \neq i} J_{ij} w_i(t)$, 
where $w_i(t)$ is the wealth of the $i$-th agent, $\eta_i(t)$ is a stochastic noise and $J$ an interaction matrix. In this case, the adjacency matrix of a network
can be included in $J$. The linear nature of the equations allows, even for a large number of agents, an effective analytical or numerical treatment. 
The drawback of the linear equations, however, is that they cannot completely describe the interaction process, since the variation of the wealth of the $i$-th agent 
depends on the wealth of the others, but not on his own wealth (apart from the stochastic term).

In a discretized kinetic theory, the state variable of a system (energy, income, $...$) is discretized into a finite number of levels or classes. 
The Boltzmann equation, which is based on the concepts of transition probabilities and detailed balance and has in general the form of an integro-differential pde, 
becomes in this approximation a system of ordinary differential equations. To obtain an accurate description of the problem, the number of classes and thus of equations and interaction parameters 
must be of the order of $10 - 100$. Then, the equations can  be quickly solved with standard numerical software. One is usually interested into the large-time behavior of the solutions. It turns out that these are asymptotic equilibrium states, which are otherwise virtually impossible to find algebraically, even with powerful symbolic software. 

The discretized Boltzmann approach represents a major progress with respect to the classical ``thermodynamical'' method based on a single rational representative agent. 
We could say that in the discretized approach there are many interacting representative agents, each one with the average features of a small subset of all individuals. 
The advantages with respect to agent-based simulations are given by a much smaller number of variables and by a ``portable'' analytical formulation independent from the software. 
The introduction of additional heterogeneity, expressed e.g. by further discrete variables beside the income, makes the model more realistic, at the price, of course, of an increase
in the number of equations.

If we want to insert a network structure into a kinetic model, we must do so in a probabilistic way, through the concept of link density $P(\alpha)$ (the fraction of nodes having $\alpha$ links) 
and of correlation functions $P(\beta | \alpha)$, which give the conditioned probability that an individual with $\alpha$ links is connected to one with $\beta$ links. 
An approach of this kind has been proposed by Boguna et al.\ \cite{Boguna}, who have described the diffusion of epidemics through differential equations giving the probability 
for each individual to be infected and in turn to infect others. It is well known that in this case the network structure is crucial, and that the hubs have a fundamental role in the transmission of the disease. 
It was found, for instance, that on certain networks there exist no minimum infectiousness threshold for the diffusion of contagion on the whole network; 
this is also the case of some viral software in the Internet. 

Our kinetic model allows the introduction of a network in a way similar as done in \cite{Boguna}, 
but with some important differences in the structure of the equations, as discussed in Sect.\ \ref{Structure}. 
This is in part because in our model for economic exchanges, money is conserved, while in a contagion process, unfortunately, the disease can multiply for free.
In Sect.\ \ref{Analytical} we establish some analytical results concerning the existence and uniqueness of solutions of the equation system and the conservation of the total wealth. In Sect.\ \ref{s4} we report the results of numerical solutions for the case of scale-free networks and compare them with those of previous works on wealth exchange models based on the Bouchaud-Mezard model and on agent-based models. Sect.\ \ref{s5} contains our conclusions.

\section{Structure of the model}
\label{Structure}

The model we discussed in \cite{BertottiModanesePhysA} is based on a system of differential equations of the kinetic discretized-Boltzmann kind. 
Society is described as an ensemble of individuals divided into a finite number of income classes; the individuals exchange money through binary and ternary interactions, 
leaving the total wealth unchanged. The interactions occur with a certain predefined frequency, 
and several other parameters can also be defined, in order to provide a probabilistic representation as realistic as possible.
For instance, we can fix the probability that in an encounter between two individuals the one who pays is the rich or the poor; 
we can make the exchanged amount depend on the income classes (variable saving propensity), etc. After a sufficiently long time the solutions of the equations reach an equilibrium state 
characterized by an income distribution, which depends on the total income and on the interaction parameters, but not on the initial distribution.

\subsection{Link and income classes, initial conditions, marginal distributions}
\label{Structure2}

As mentioned in the Introduction, we would like to introduce now in our model an additional important element of heterogeneity of the population: 
for this purpose, we suppose that different individuals have in general a different number 
of links of economic nature with other individuals. Let this number of links vary between 1 and a maximum value $N$. The population will be divided into $N$ subsets, 
each one comprising the individuals which have the same number of links. This subdivision is regarded as fixed during the evolution, in the sense that 
each individual maintains the same number of links when it becomes richer or poorer following the interaction with others. The assumption of constant link distribution 
is clearly only an approximation to a real situation. It can be regarded as adequate for situations where the evolution of the income distribution is ``soft'' 
in the sense that it does not involve massive migrations between classes, which would also likely  imply significant changes in the connection network.

When we further subdivide the population into $n$ income classes as usual, the income distribution of the society at a given instant $t$ is defined 
by the densities $x^\alpha_i(t)$, with $\alpha = 1, \dots ,N$ and $i = 1, \dots ,n$. The function $x^\alpha_i(t)$ represents the fraction of individuals 
which belong to the $i$-th income class and have $\alpha$ economic links to other individuals of the population. As we will show in  Sect.\ \ref{Analytical},
the densities $x^\alpha_i(t)$ are normalized at all times: $\sum_{\alpha, i} x^\alpha_i(t)=1$. 
We  also consider the marginal densities $x^\alpha (t)= \sum_{ i} x^\alpha_i(t)$ and $x_i(t)=\sum_{\alpha} x^\alpha_i(t)$.

The average income of the society (which, due to the normalization of the population, also coincides 
with the total income) is
\begin{equation}
\mu  = \sum\limits_{\alpha  = 1}^N {\sum\limits_{i = 1}^n  {r_i}x_i^\alpha } (t) \ ,
\end{equation}
where $r_i$ is the income of the $i$-th class. As will be shown in Sect.\ \ref{Analytical}, this is a conserved quantity, constant during the time evolution. The average income of the individuals with $\alpha$ links is
\begin{equation}
\mu^\alpha(t)  = {\sum\limits_{i = 1}^n  {r_i}x_i^\alpha } (t) \ .
\end{equation}

Obviously $\mu  = \sum\limits_{\alpha  = 1}^N  \mu^\alpha (t)$, but while $\mu$ is a constant,  the $\alpha$-averages $\mu^\alpha (t)$ change in general in time.
Indeed, money circulates among the link classes. It is interesting to compare the incomes $\mu^\alpha (t)$ at equilibrium, 
in order to see which link classes $\alpha$ comprise the individuals who are, in the average, the richest or the poorest. It is not clear in advance 
whether a large number of links leads, in general, to a larger average income.

The simplest way to fix the initial conditions is to assume that all individuals are at $t=0$ in the same income class. Previous experience with the model indicates that, 
for instance, if $n=25$ and the class incomes $r_i$ grow linearly, then starting off the evolution with all the individuals in a class between the 5th and the10th 
typically allows a smooth convergence towards a unique 
equilibrium distribution depending only on the total income 
and exhibiting a fat tail. 

Suppose then that $x_i^\alpha(0)=0$ for $i \neq i_0$, where $i_0$ is properly chosen as mentioned above. 
According to our previous definitions, we must set $x_{i_0}^\alpha(0)=P(\alpha)$, i.e.\ we distribute the individuals at the beginning 
according to their link classes, all their incomes being initially equal. 

More generally, if not all individuals are initially in the same income class, then the condition $\sum_i x_i^\alpha(0)=P(\alpha)$ must be satisfied. 

\subsection{Evolution equations}

In this paper, for simplicity, we shall incorporate a network structure in a particular version of our model, 
namely a version without tax payment and redistribution. The corresponding evolution equations describe only binary exchanges and take the form
\begin{equation}
\frac{{d{x_i}(t)}}{{dt}} = \sum\limits_{h,k = 1}^n {C_{hk}^i{x_h}(t){x_k}(t)} - 
\sum\limits_{h,k = 1}^n {C_{ik}^h{x_i}(t){x_k}(t)} \ ,
\end{equation}
where the constant coefficients $C_{hk}^i$, satisfying for any fixed $h$ and $k$
the condition $\sum_{i=1}^n C_{hk}^i = 1$, 
express the probability 
that an individual of the $h$-th class will belong to the 
$i$-th class after a direct interaction with an individual of the $k$-th class;
they define 
all the features of the model, as described in detail in \cite{BertottiModanesePhysA}.

We can generalize the evolution equations for the densities $x_i^\alpha(t)$ by introducing the network structure as follows:

\begin{equation}
\frac{{dx_i^\alpha (t)}}{{dt}} = 
\sum\limits_{h,k = 1}^n {\sum\limits_{\beta  = 1}^N  C_{h, \alpha, k, \beta}^{i, \alpha} \, x_h^\alpha (t) x_k^\beta (t)} 
- 
\sum\limits_{h,k = 1}^n {\sum\limits_{\beta  = 1}^N  C_{i, \alpha, k, \beta}^{h, \alpha} \, x_i^\alpha (t) x_k^\beta (t)} 
\ .
\label{evolution}
\end{equation}




The coefficients $C_{h, \alpha, k, \beta}^{i, \alpha} \in [0,1]$ in  (\ref{evolution}) express the 
probability 
that an individual of the $h$-th class, with $\alpha$ links
will belong to the 
$i$-th class (maintaining the same number $\alpha$ of links) after an interaction with an
individual of the $k$-th class, with $\beta$ links.
They have to satisfy
\begin{equation}
\sum_{i=1}^n C_{h, \alpha, k, \beta}^{i, \alpha} = 1 \ ,
\label{c=1}
\end{equation}
for any fixed $h, k, \alpha, \beta$.
We take them here as
\begin{equation}
C_{h, \alpha, k, \beta}^{i, \alpha} = a_{h, \alpha, k, \beta}^{i, \alpha} + b_{h, \alpha, k, \beta}^{i, \alpha} \ ,
\label{c}
\end{equation}
where 
the only nonzero elements
$a_{h, \alpha, k, \beta}^{i, \alpha}$ are those for which $h=i$, i.e. are
\begin{equation}
a_{i, \alpha, k, \beta}^{i, \alpha} = 1 \ ,
\label{a}
\end{equation}
and
the only possibly nonzero elements $b_{h, \alpha, k, \beta}^{i, \alpha}$ are
those of the form 
\begin{eqnarray}
b_{i+1,\alpha, k, \beta}^{i, \alpha}& = 
                  & p_{i+1,k} \, \frac{S}{r_{i+1} - r_{i}} \, P(\beta | \alpha) \ , \nonumber \\
b_{i,\alpha, k, \beta}^{i, \alpha} & = 
            & - \, p_{k,i} \, \frac{S}{r_{i+1} - r_{i}} \, P(\alpha | \beta)
               - \, p_{i,k} \, \frac{S}{r_{i} - r_{i-1}} \, P(\beta | \alpha) \ ,  \nonumber \\
b_{i-1,\alpha, k, \beta}^{i, \alpha} & = 
               & p_{k,i-1} \, \frac{S}{r_{i} - r_{i-1}} \, P(\alpha | \beta) \ . 
\label{b}
\end{eqnarray} 
Concerning the terms in $(\ref{b})$, 

\begin{itemize} 
\item[-] we notice first of all that
the first line is meaningful only for $i \le n-1$ and $k\le n-1$
and the third line only for $i \ge 2$ and $k\ge 2$; as for the second line, 
the first addendum on the r.h.s. is effectively present only provided $i \le n-1$ and $k \ge 2$ and the second addendum only provided $i \ge 2$ and $k \le n-1$.
\item[-] The coefficients $p_{h,k} \in [0,1]$ for $h, k = 1, ..., n$
express the probability that  in an encounter between an individual of the $h$th-income class and an individual of the $k$th-one,
the one who pays is the $h$-individual. As discussed in \cite{BertottiModaneseEPJB}, they carry information, in particular, on the level of heterogeneity of individuals
belonging to different classes.
To fix the ideas, we take them here as
$$
p_{h,k} = \min \{r_h,r_k\}/{4 r_n} \ ,
$$
with the exception of the terms
$p_{j,j} = {r_j}/{2 r_n}$ for $j = 2, ..., n-1$,
$p_{h,1} = {r_1}/{2 r_n}$ for $h = 2, ..., n$, 
$p_{n,k} = {r_k}/{2 r_n}$ for $k = 1, ..., n-1$,
$p_{1,k} = 0$ for $k = 1, ..., n$
and
$p_{hn} = 0$ for $h = 1, ..., n$. 
\item[-] $S$ denotes the amount of money which is exchanged in an interaction.
\item[-] The second-order correlation function $P\left( {\beta |\alpha } \right)$ denotes the probability that an individual with $\alpha$ links is connected to an individual with $\beta$ links. 
It is normalized as $\sum_\beta P\left( {\beta |\alpha } \right)=1$ and it is a feature of the chosen network structure. It is 
related to the link density $P(\alpha)$ by a ``closure condition'' \cite{correlations}, which has the form
\begin{equation}
\alpha P(\beta|\alpha) P(\alpha) = \beta P(\alpha|\beta) P(\beta), \qquad \forall \alpha ,\beta  = 1, \ldots ,N \ .
\label{cnc}
\end{equation}
\end{itemize}
All this guarantees that the condition $(\ref{c=1})$ is verified.


Note that in the equations used for the epidemic model of \cite{Boguna} an individual with $\alpha$ links is, 
compared with an individual with one link, $\alpha$ times more likely to contract the infection. This is a natural assumption in the case of an epidemic, 
and highlights the role of hubs in the diffusion of the disease. In our model of economic interactions it is possible to introduce  in (\ref{b}) suitable ``amplification'' factors, say $f_\alpha$, so that 
an individual with $\alpha$ links has a volume of economic exchanges $f_\alpha$ times larger than an individual with one link. This could be regarded 
as a reasonable assumption or not, 
depending on the example which we have in mind and on the context in which we intend to apply the model. If we think of the case of the owners of a firm or a store, 
we conclude that in that case more links mean a larger total volume of economic exchanges. In other cases, however, the model without amplification 
is more realistic, implying that an individual with $\alpha$ links has the same total economic exchange as an individual with one link, but distributed 
among the various links. This could correspond to the case of a professional who furnishes some service and divides his working time among several clients.

\section{Some analytical results}
\label{Analytical}


We prove here that for the eq.s (\ref{evolution}) the existence and uniqueness for all $t \geq 0$ of the solution with prescribed initial conditions hold true. Furthermore, we prove that the total wealth remains constant during the evolution.

Denote 
\begin{equation}
\Sigma = \{ x = \{x^\alpha_i\}, \ \hbox{such that} \ x^\alpha_i \ge 0 \ \forall i=1, ..., n, \alpha=1, ..., N \ \hbox{and} \ \sum_{i=1}^n \sum_{\alpha =1}^N  x^\alpha_i = 1\} \ .
\end{equation}

\noindent {\bf Theorem 1} 
Let the elements $C_{h, \alpha, k, \beta}^{i, \alpha}$
(with $i, h, k = 1, ..., n$, $\alpha, \beta = 1, ..., N$)
be as in $(\ref{c}), (\ref{a}), (\ref{b})$.
For any $x(0) \in \Sigma$,
the solution $x(t) = \{x^\alpha_i (t)\}$ of the Cauchy problem for the equations $(\ref{evolution})$
with initial condition $x(0)$ exists and is unique for all $t \in [0,+\infty)$. Moreover, 
$\forall t \geq 0$ one has
\begin{equation}
x^\alpha_i (t) \ge 0 \quad
\forall i = 1, \ldots , n, \forall \alpha = 1, \dots , N \qquad \hbox{and} \qquad \sum_{i=1}^n \sum_{\alpha =1}^N x^\alpha_i (t) = 1 \ .
\label{solution in the future}
\end{equation}
\label{theorem: well-posedness}

\noindent {\bf Proof:} The r.h.s. of the equations $(\ref{evolution})$ is of class $C^{\infty}$ and, in particular, locally Lipschitzian. This guarantees the 
local existence of a unique solution $x(t) = \{x^\alpha_i (t)\}$ of the Cauchy problem. Then, one has
$x_i^\alpha (t) = x_i^\alpha (0) + \int_0^t \, \frac{{dx_i^\alpha}}{{ds}}\, ds$ and 
\begin{eqnarray}
& & \sum_{i=1}^n \sum_{\alpha =1}^N x^\alpha_i (t) = \sum_{i=1}^n \sum_{\alpha =1}^N x^\alpha_i (0) \nonumber \\
&+&
\int_0^t \, 
\bigg [
\sum\limits_{i,h,k = 1}^n  \sum\limits_{\alpha,\beta  = 1}^N  C_{h, \alpha, k, \beta}^{i, \alpha} \, x_h^\alpha (s) x_k^\beta (s)
- 
\sum\limits_{i,h,k = 1}^n \sum\limits_{\alpha,\beta  = 1}^N  C_{i, \alpha, k, \beta}^{h, \alpha} \, x_i^\alpha (s) x_k^\beta (s)
\bigg ]
\, ds = 1 
\end{eqnarray}
as well, for any $t \ge 0$, for which $x_i^\alpha (t)$ exists. This ensures that $\sum_{i=1}^n \sum_{\alpha =1}^N x^\alpha_i (t) = 1$ as long as the solution $x(t)$ exists.

We denote now
$$
G_i^{\alpha} (x(s)) = \sum\limits_{h,k = 1}^n  \sum\limits_{\beta  = 1}^N  C_{h, \alpha, k, \beta}^{i, \alpha} \, x_h^\alpha (s) x_k^\beta (s) \ .
$$
Then, also in view of (\ref{c=1}), the equations (\ref{evolution}) can be written as
\begin{equation}
\frac{{dx_i^\alpha (t)}}{{dt}} = 
G_i^{\alpha} (x(t))
- 
x_i^\alpha (t) 
\ ,
\label{evolutionsimple}
\end{equation}
which implies that, if $x_i^\alpha (t)$ is a solution of $(\ref{evolutionsimple})$, 
$\frac{{d}}{{dt}} e^{t} x_i^\alpha (t) = e^{t} G_i^{\alpha} (x(t))$. Consequently,
$$
x_i^\alpha (t) = e^{-t} x_i^\alpha (0) + \int_0^t \, e^{- t + s} G_i^{\alpha} (x(s)) \, ds \ ,
$$
and this proves that $x^\alpha_i \ge 0$ for $i=1, ..., n$ and $\alpha=1, ..., N$, for any $t \ge 0$, for which $x_i^\alpha (t)$ exists. 

Since $x(t)=\{x^\alpha_i (t)\}$ remains within a compact region in ${\bf R}^n$, its existence for all $t \ge 0$ holds true.
\QED

\smallskip

We show next that the
total wealth expressed by the function 
\begin{equation}
W : \Sigma \to {\bf R} \ , \qquad W(x) = \sum_{i=1}^n \sum_{\alpha =1}^N r_i x^\alpha_i 
\label{total wealth}
\end{equation}
is a conserved quantity. Indeed we prove the following result.

\noindent {\bf Theorem 2} 
The value of the total wealth $(\ref{total wealth})$ remains constant along any solution of $(\ref{evolution})$.
\label{theorem: wealth conservatively}

\noindent {\bf Proof:} It follows from $(\ref{c=1}), (\ref{c}), (\ref{a}), (\ref{b})$ and $(\ref{solution in the future})$ that
\begin{eqnarray}
{{d W} \over {d t}} & = &  
\sum_{i=1}^n \sum_{\alpha =1}^N r_i \frac{{dx_i^\alpha}}{{dt}}  \nonumber \\
& =  & 
\sumsui \sumsua \sumsuk \sumsub r_i x^\alpha_i x^\beta_k 
+
\sumsui \sumsua \sumsuh \sumsuk \sumsub r_i b_{h, \alpha, k, \beta}^{i, \alpha} x^\alpha_h x^\beta_k
-
\sumsui \sumsua r_i x^\alpha_i \nonumber \\
& =  & 
\sum_{h=2}^{n-1} \sum_{k=2}^{n-1} \sumsui \sumsua \sumsub     r_i b_{h, \alpha, k, \beta}^{i, \alpha} x^\alpha_h x^\beta_k
\nonumber \\
& & +
\sumsuk \sumsui \sumsua \sumsub     r_i b_{1, \alpha, k, \beta}^{i, \alpha} x^\alpha_1 x^\beta_k
+
\sumsuk \sumsui \sumsua \sumsub     r_i b_{n, \alpha, k, \beta}^{i, \alpha} x^\alpha_n x^\beta_k
\nonumber \\
& & +
\sum_{h=2}^{n-1} \sumsui \sumsua \sumsub     r_i b_{h, \alpha, 1, \beta}^{i, \alpha} x^\alpha_h x^\beta_1
+
\sum_{h=2}^{n-1} \sumsui \sumsua \sumsub     r_i b_{h, \alpha, n, \beta}^{i, \alpha} x^\alpha_h x^\beta_n
\label{Lie derivative}
\end{eqnarray} 
We will now argue separately on the terms on the r.h.s. in $(\ref{Lie derivative})$.

$\bullet$ To start with, we notice that, if $2 \le h \le n-1$ and $2 \le k \le n-1$, then
\begin{eqnarray}
\sumsui r_i b_{h, \alpha, k, \beta}^{i, \alpha}
&=&
r_{h-1} b_{h, \alpha, k, \beta}^{h-1, \alpha}   +   r_h b_{h, \alpha, k, \beta}^{h, \alpha}   +   r_i b_{{h+1}, \alpha, k, \beta}^{h+1, \alpha} \nonumber \\
&=&
p_{k,h} \, S \, P(\alpha | \beta) - p_{h,k} \, S \, P(\beta | \alpha) \nonumber \ .
\end{eqnarray} 
Equivalently, the quantity $Q_{h,k}^{\alpha,\beta} = p_{k,h} \, S \, P(\alpha | \beta) - p_{h,k} \, S \, P(\beta | \alpha)$ is antisymmetric.
This implies that 
\begin{equation}
\sum_{h=2}^{n-1} \sum_{k=2}^{n-1} \sumsui \sumsua \sumsub     r_i b_{h, \alpha, k, \beta}^{i, \alpha} x^\alpha_h x^\beta_k
=
\sum_{h=2}^{n-1} \sum_{k=2}^{n-1} \sumsua \sumsub     Q_{h,k}^{\alpha,\beta} x^\alpha_h x^\beta_k
=
0 \ ,
\end{equation}
i.e., the first term on the r.h.s. in $(\ref{Lie derivative})$ is zero.

$\bullet$ Straightforward calculations show that

\begin{itemize}
\item[-] the second term on the r.h.s. in $(\ref{Lie derivative})$ is equal to
\begin{equation}
\sumsua \sumsub \sumsuk \big(r_1 b_{1, \alpha, k, \beta}^{1, \alpha} x^\alpha_1 x^\beta_k+r_2 b_{1, \alpha, k, \beta}^{2, \alpha} x^\alpha_1 x^\beta_k\big)
=
\sumsua \sumsub \sum_{k=2}^{n} p_{k,1} \, S \, P(\alpha | \beta) x^\alpha_1 x^\beta_k \ .
\label{2 term}
\end{equation}
\item[-] the third term on the r.h.s. in $(\ref{Lie derivative})$ is equal to
\begin{equation}
\sumsua \sumsub \sumsuk \big(r_{n-1} b_{n, \alpha, k, \beta}^{n-1, \alpha} x^\alpha_n x^\beta_k+r_n b_{n, \alpha, k, \beta}^{n, \alpha} x^\alpha_n x^\beta_k\big)
=
\sumsua \sumsub \sum_{k=1}^{n-1} - p_{n,k} \, S \, P(\beta | \alpha) x^\alpha_n x^\beta_k \ .
\end{equation}
\item[-] the fourth term on the r.h.s. in $(\ref{Lie derivative})$ is equal to
\begin{equation}
\sum_{h=2}^{n-1} \sumsua \sumsub     \big(r_{h-1} b_{h, \alpha, 1, \beta}^{{h-1}, \alpha} x^\alpha_h x^\beta_1+r_h b_{h, \alpha, 1, \beta}^{h, \alpha} x^\alpha_h x^\beta_1\big)
=
\sumsua \sumsub \sum_{h=2}^{n-1} - p_{h,1} \, S \, P(\beta | \alpha) x^\alpha_h x^\beta_1 \ .
\end{equation}
\item[-] the fifth term on the r.h.s. in $(\ref{Lie derivative})$ is equal to
\begin{equation}
\sum_{h=2}^{n-1} \sumsua \sumsub     \big(r_h b_{h, \alpha, n, \beta}^{h, \alpha} x^\alpha_h x^\beta_n+r_{h+1} b_{h, \alpha, n, \beta}^{{h+1}, \alpha} x^\alpha_h x^\beta_n\big)
=
\sumsua \sumsub \sum_{h=2}^{n-1} p_{n,h} \, S \, P(\alpha | \beta) x^\alpha_h x^\beta_n \ .
\label{4 term}
\end{equation}
\end{itemize}

Then, adding the four terms in $(\ref{2 term}) - (\ref{4 term})$,
we get
$$
\sumsua \sumsub p_{n,1} \, S \, P(\alpha | \beta) x^\alpha_1 x^\beta_n - \sumsua \sumsub p_{n,1} \, S \, P(\alpha | \beta) x^\alpha_1 x^\beta_n = 0 \ .
$$
Namely, 
the sum of the terms from the second one to the fifth one on the r.h.s. in $(\ref{Lie derivative})$,
is zero.

Summarizing, the r.h.s. in $(\ref{Lie derivative})$ is equal to zero, which proves the claim.
\QED

\section{Numerical results. Discussion}
\label{s4}

We have solved the equations numerically in some simple low-dimensional cases, varying the initial conditions and the features of the network. More precisely, we have considered scale-free networks with link density $P(\alpha)=c/\alpha^q$, varying the exponent $q$. We have also varied the correlation matrices $P(\beta|\alpha)$, focusing on the cases of assortative and disassortative correlations, but we found that this has little influence on the equilibrium distributions.

Concerning the dependence on the initial conditions, the results show that the equilibrium distributions $\hat x_i^\alpha \equiv x_i^\alpha(+\infty)$ all share the following significant properties.

\begin{enumerate}

\item
The marginal income equilibrium distribution $\hat x_i=\sum_{\alpha} x^\alpha_i(+\infty)$ depends only on the total income $\mu$.

\item
The detailed income equilibrium distribution $\hat x_i^\alpha$ depends on $\mu$ and on $x^\alpha(0)$, but not on the detailed initial condition $x_i^\alpha(0)$.

\end{enumerate}

Further interesting results concern the difference between the histograms of the quantities $\{x_i^1,i=1\ldots n\}, \{x_i^2,i=1\ldots n\}, \ldots ,\{x_i^N,i=1\ldots n\}$ which represent the income distributions of the various ``link classes'', i.e.\ of the classes comprising individuals with $1, 2, \ldots N$ links. As mentioned in the Introduction, one could expect that more connected individuals generally tend to acquire more wealth. (The strength of this effect is also expected to depend on the possible presence of an ``amplified'' coupling, compare Sect.\ \ref{Structure}.) It turns out, however, that the relative wealth of the link classes is determined by the exponent $q$ of the power law of the link density $P(\alpha)=c/\alpha^q$, the critical value being $q=1$. For $q<1$ the rich individuals tend to concentrate in the classes with many links, for $q>1$ in those with few links.

\begin{figure}
\begin{center}
  \includegraphics[width=7cm,height=3.5cm]{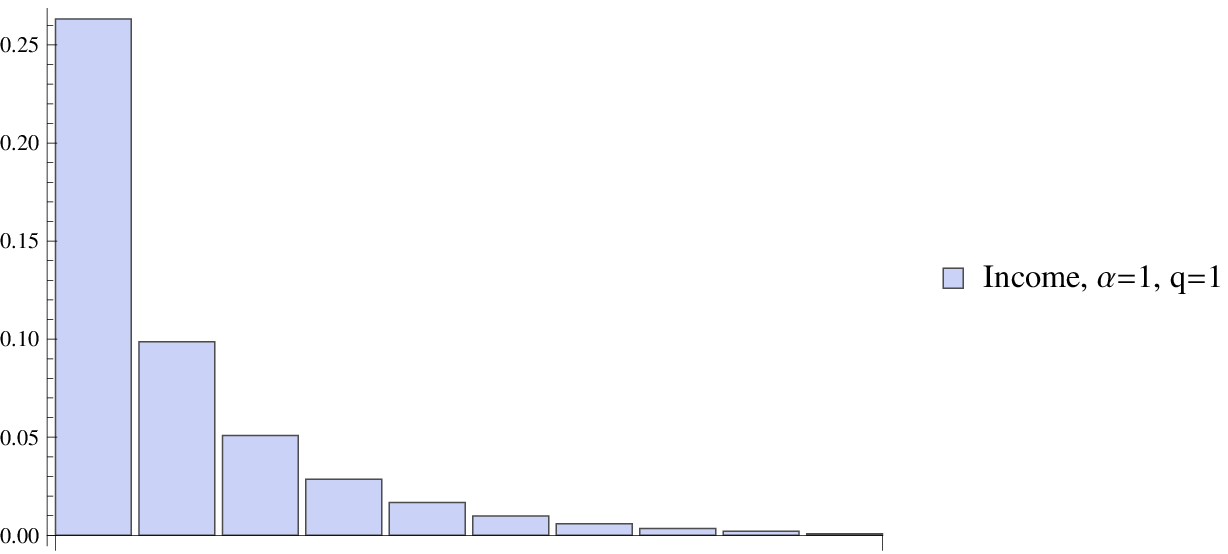}
   \hskip0.7cm
  \includegraphics[width=7cm,height=3.5cm]{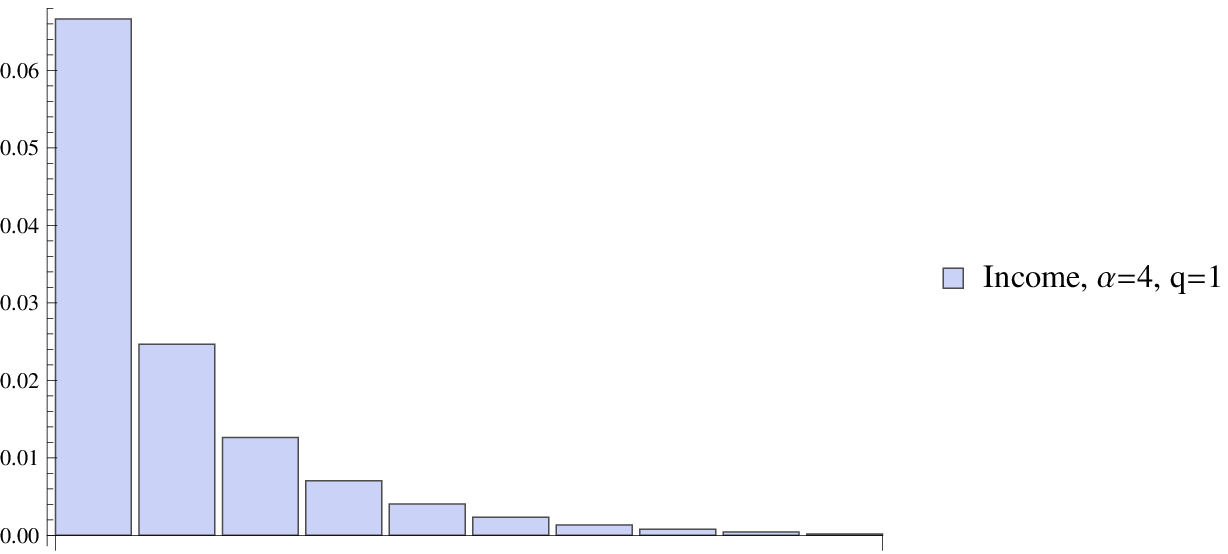}
\caption{Income histograms for link density $P(\alpha)=c/\alpha^q$ with $q=1$. The total population has been divided into 10 income classes and 4 link classes. Left: populations of the 10 income classes of individuals with 1 link. Right: the same for individuals with 4 links. The histograms of the link classes are all geometrically similar. This reminds the energy distribution of a gas containing molecules of different kinds, all in equilibrium at the same temperature. The corresponding Gini indices (Table \ref{Gini1}) are all equal. } 
\label{pippo}
\end{center}  
\end{figure}

When $P(\alpha)=c/\alpha$ the histograms of the link classes are all similar (Fig.\ \ref{pippo}), and more exactly one has for any $i$, $j$, $\alpha$, $\beta$
\begin{equation}
\frac{\hat x^\alpha_i}{\hat x^\alpha_j} = \frac{\hat x^\beta_i}{\hat x^\beta_j}
\end{equation}
or equivalently
\begin{equation}
\frac{\hat x^\alpha_i}{\hat x^\beta_i} = \frac{\hat x^\alpha_j}{\hat x^\beta_j}
\end{equation}

This reminds the energy distribution of a gas containing molecules of different kinds, all in equilibrium at the same temperature. In this analogy, individuals with more links correspond to molecules with larger cross-section. 

\begin{figure}
\begin{center}
  \includegraphics[width=7cm,height=3.5cm]{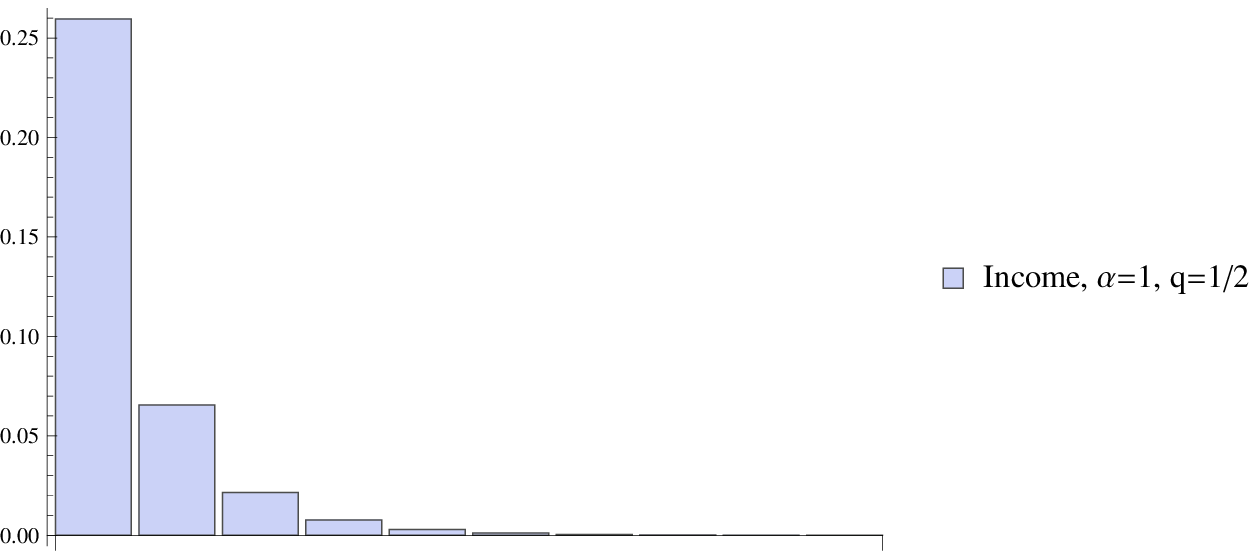}
   \hskip0.7cm
  \includegraphics[width=7cm,height=3.5cm]{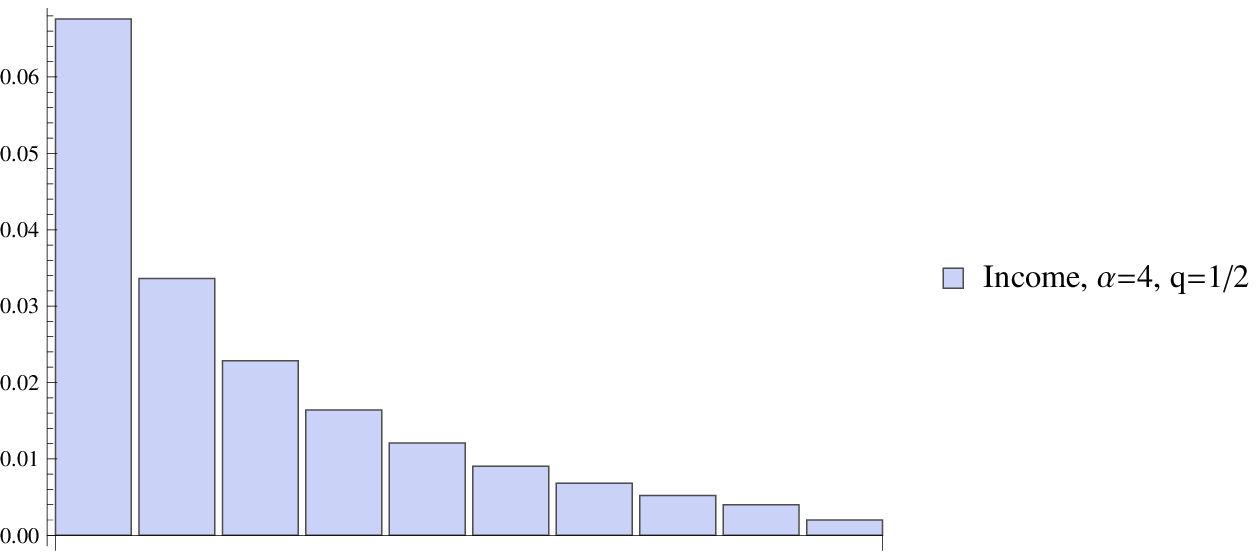}
\caption{Income histograms for link density $P(\alpha)=c/\alpha^q$ with $q=1/2$. The total population has been divided into 10 income classes and 4 link classes. Left: populations of the 10 income classes of individuals with 1 link. Right: the same for individuals with 4 links. The fraction of wealthy individuals is clearly much larger in the second case. For $q=2$, however, the relation is reversed (Fig.\ \ref{f1bis}) and for $q=1$ the two istograms have exactly the same form (Fig.\ \ref{pippo}). } 
\label{f1}
\end{center}  
\end{figure}

When $P(\alpha)=c/\alpha^q$ with $q<1$, the histograms of the classes with few links display a concentration of individuals in the lowest income classes (Fig.\ \ref{f1}). The corresponding Gini indices $G^\alpha$ (Table \ref{Gini1}) are low for small $\alpha$, because this concentration amounts to smaller inequality; in simple terms, in this case the less connected people all tend to be poor. For $q>1$, on the contrary, it is the most connected people that tend to concentrate in the lowest income classes. This appears to mean that when the hubs in the network are very few, the mechanism of purely kinetic wealth exchange puts them at a disadvantage. These results emerge very clearly from the numerical solutions, although we did not consider a large maximum number $N$ of links yet, since it is not trivial in general to write for any $q$ large correlation matrices $P(\beta|\alpha)$ which satisfy the network closure condition (\ref{cnc}).

We have mainly considered scale-free networks in our numerical solutions, because there is ample empirical evidence that they are dominant in socio-economic interactions. This is confirmed also by theoretical models in which the network evolves dynamically during the interactions \cite{Korean}. We are not discussing dynamical networks here for two reasons: first, we intend to use our model to simulate the effects of fiscal, welfare and macro-economic policies on systems which are close to equilibrium, and not in phases of strong growth or economic collapse, where the network structure undergoes radical changes; second, we would like to compare our results with similar econophysics models, namely the Bouchaud-Mezard model and the Yard Sale model, whose network versions also run on a fixed network structure \cite{Mexican,Loffredo2}. In models with dynamical network structure \cite{Abram,Iglesias,Coelho,Hu,Korean}, it is often observed that highly connected  individuals acquire more wealth, and conversely rich individuals get more and more connected. A direct comparison to our case, however, is very difficult, because in \cite{Abram,Iglesias,Coelho,Hu,Korean} the rules for the growth of the network are strongly model-dependent, and the rules for the economic exchanges contain decisional and game-theoretical elements which can be simulated only in an agent-based version.

Finally note that for fixed link density $P(\alpha)$, the dependence of the behavior described above on the assortative or dissortative nature of the matrices $P(\beta|\alpha)$ is very weak. 

These results are compatible with those of the cited models \cite{Mexican,Loffredo2}. Bustos-Guajardo and Moukarzel \cite{Mexican} find (like us) only a weak dependence of the total income distribution on the network, but observe a strong dependence on the network of the relaxation time, because the network structure has an important effect on the diffusion of wealth in time. We also see similar effects in the approach of the numerical solutions to equilibrium (Fig.\ 2). For $q>1$ the classes with more links tend to lose wealth, starting from equal initial conditions, in favor of those with few links. For $q<1$ the opposite happens, and for $q=1$ there is essentially no transfer of wealth between link classes. Loffredo and Garlaschelli \cite{Loffredo2} find that in the case of spatially homogenous networks, like those we considered, the total income distribution is log-normal; our total distributions are also compatible with a log-normal curve. Loffredo and Garlaschelli  further observe a transition to a log-normal plus power tail in the presence of higher-order correlations, which were not explored in this paper.

\begin{table}[h]
\begin{center}
\begin{tabular}{|c|c|c|c|}
\hline
Link class $\alpha$  & \ Gini, $q=1/2$  \ & \ Gini, $q=1$  \ & \ Gini, $q=2$ \ \nonumber \\
\hline
1 & \ 0.24  \ & \ 0.36 \ & \ 0.38 \nonumber \\
\hline
2 & \ 0.33 \ & \ 0.36 \ & \ 0.21 \nonumber \\
\hline
3 & \ 0.38 \ & \  0.35 \ & \ 0.14 \nonumber \\
\hline
4 & \ 0.40 \ & \  0.35 \ & \ 0.10 \nonumber \\
\hline
Tot.\ popul. & \ 0.37 \ & \ 0.36 \ & \ 0.37 \nonumber \\
\hline
\end{tabular}
\end{center}
\caption
{Gini indices of the income histograms of the classes with $\alpha$ links and of the total population, in the case of scale-free networks with link density $P(\alpha)=c/\alpha^q$.}
\label{Gini1}
\end{table}

\section{Conclusion}
\label{s5}

In this work we have set up a consistent mathematical framework, based on kinetic differential equations, for the investigation of economic exchanges between individuals linked by a fixed network of economic relations. The introduction of a network structure has been previously obtained for agent-based models and for the Bouchaud-Mezard stochastic theory; it is a novelty in kinetic theory. The use of the network clearly allows a better representation of real situations and is an important way of introducing heterogeneity into the model. Our results, summarized in Sect.\ \ref{s4} are consistent with some previous results. In addition, our formalism allows us to examine the incomes of individuals having a different number of links and to test the idea that more connected people tend to become richer. We find that for scale-free networks with link density $P(\alpha)=c/\alpha^q$ this is true only if $q<1$. A direct comparison with models of economic exchanges which include dynamical, disordered or directed networks is not possible. Our next objective is not to consider the evolution of the network, but to study the effect on a pre-existent network of different economic policies, in line with our previous works where the effects of taxation, tax evasion and welfare measures have been explored. For that purpose, we will introduce in future work the redistribution terms; we decided to omit them here mainly because they would have made the analytical proofs given in this paper exceedingly complicated;  the main interaction mechanism, however, is already contained in the quadratic terms.

\vskip 10 cm

\begin{figure}
\begin{center}
  \includegraphics[width=8cm,height=6cm]{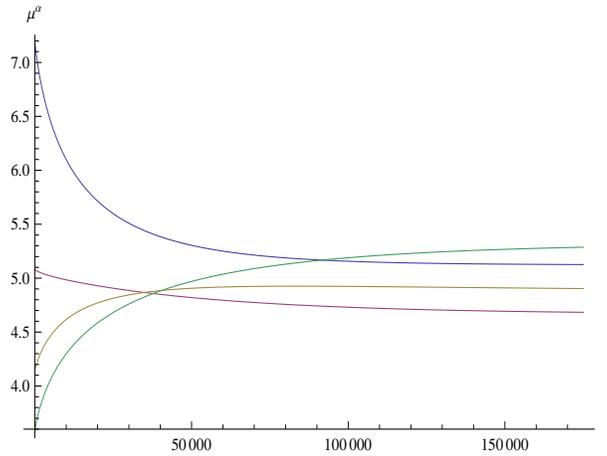}
\caption{Time evolution of the total incomes $\mu^\alpha$ of individuals with $\alpha$ links in the case of link density $P(\alpha)=c/\alpha^q$, with $q=1/2$, $\alpha =1,2,3,4$ ($\alpha = 1$: upper curve). At the initial time, individuals with few links have larger total income because they are more numerous, but then they lose wealth in favor of those with more links.
Consider, in particular, the individuals with 1 and 4 links, whose detailed equilibrium income distribution is represented in the histograms of Fig.\ \ref{f1}. Since $P(1)=2P(4)$, the number of individuals with 1 link is fixed to be twice as large as that of individuals with 4 links. At the initial time, their total income is also twice as large, because all individuals start in the same income class. At equilibrium, however, the total income of the individuals with 1 link has become smaller than the total income of the individuals with 4 links. This means that the average income of the least connected individuals is  now less than half the average income of the most connected individuals. More exactly, some of the most connected individuals has become very rich (compare Fig.\ \ref{f1}), building up a tail in the income distribution, while all the individuals with 1 link have got concentrated in the low-income classes. For $q=2$, exactly the opposite happens (compare Fig.\ \ref{f2bis}).} 
\label{f2}     
\end{center}
 \end{figure}

\begin{figure}
\begin{center}
  \includegraphics[width=7.5cm,height=3.5cm]{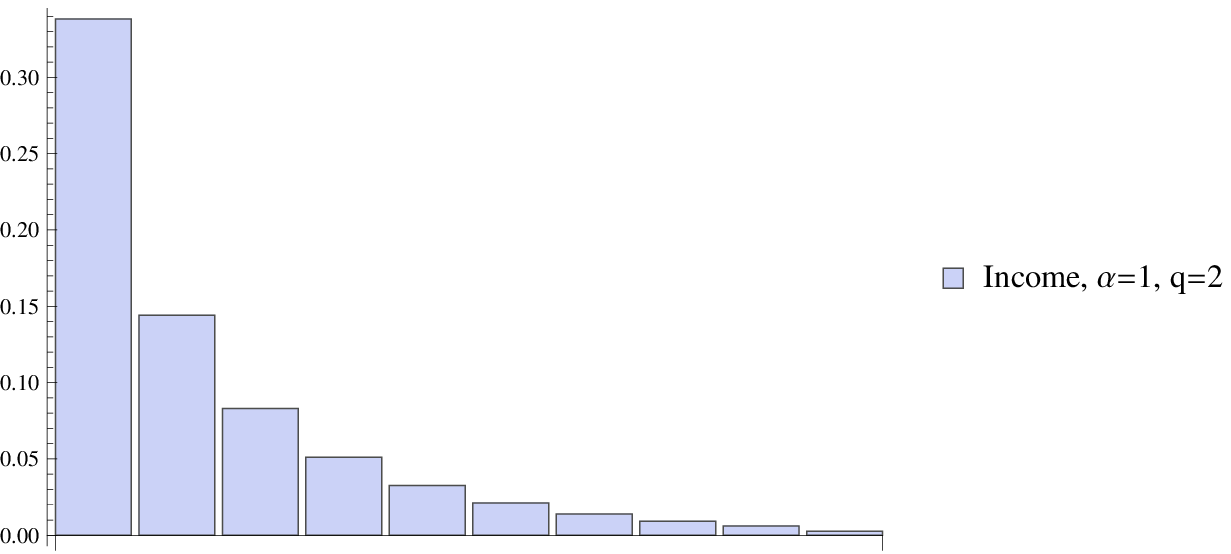}
   \hskip0.7cm
  \includegraphics[width=7.5cm,height=3.5cm]{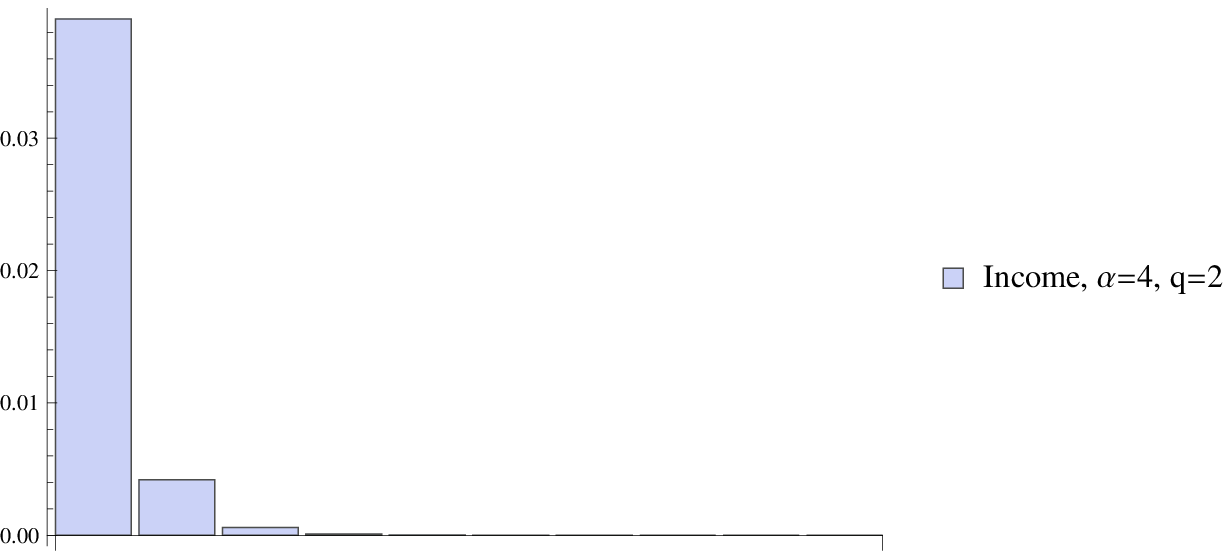}
\caption{Income histograms for link density $P(\alpha)=c/\alpha^q$ with $q=2$. The total population has been divided into 10 income classes and 4 link classes. Left: populations of the 10 income classes of individuals with 1 link. Right: the same for individuals with 4 links. The fraction of wealthy individuals is clearly much larger in the first case. Compare Fig.\ \ref{f1}. } 
\label{f1bis}   
\end{center}
\end{figure}

\begin{figure}
\begin{center}
  \includegraphics[width=8cm,height=6cm]{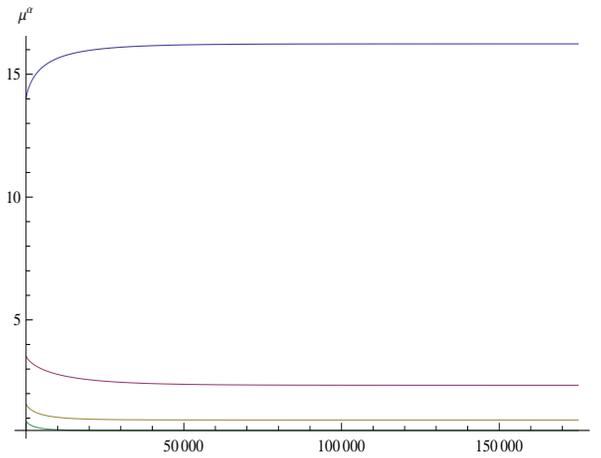}
\caption{Time evolution of the total incomes $\mu^\alpha$ of individuals with $\alpha$ links in the case of link density $P(\alpha)=c/\alpha^q$ with $q=2$. At the initial time, individuals with few links have larger total income because they are more numerous; then those with 1 link gain further wealth from those with more links. Compare Fig.\ \ref{f2}.} 
\label{f2bis}   
\end{center}   
\end{figure}

\end{document}